\documentstyle[aps,prc,epsfig]{revtex}

\begin{document}
\draft
\title{$p_T-$fluctuations in high-energy p-p and A-A collisions}

\author{Robert Korus\footnote{Electronic address: 
{\tt korus@pu.kielce.pl}}}

\address{Institute of Physics, \'Swi\c etokrzyska Academy, \\
ul. Konopnickiej 15, PL - 25-406 Kielce, Poland}

\author{Stanis\l aw Mr\' owczy\' nski\footnote{Electronic address:
{\tt mrow@fuw.edu.pl}}}

\address{So\l tan Institute for Nuclear Studies, \\
ul. Ho\.za 69, PL - 00-681 Warsaw, Poland \\
and Institute of Physics, \'Swi\c etokrzyska Academy, \\
ul. Konopnickiej 15, PL - 25-406 Kielce, Poland}

\date{3-rd July 2001}

\maketitle

\begin{abstract}

The event-by-event $p_T-$fluctuations in proton-proton and central 
Pb-Pb collisions, which have been experimentally studied by means 
of the so-called $\Phi-$measure, are analyzed. The contribution due to 
the correlation which couples the average $p_T$ to the event multiplicity is 
computed. The correlation appears to be far too weak to explain the 
preliminary experimental value of $\Phi (p_T)$ in p-p interactions. 
The significance of the result is discussed.

\end{abstract}

\vspace{0.5cm}

 

\section{Introduction}

The transverse momentum fluctuations in proton-proton and central Pb-Pb 
collisions at 158 GeV per nucleon have been recently measured \cite{App99}
on event-by-event basis. To eliminate trivial `geometrical' fluctuations due 
to the impact parameter variation the so-called $\Phi-$measure \cite{Gaz92} 
has been used. $\Phi$ is constructed is such a way that it is exactly the same 
for nucleon-nucleon (N-N) and nucleus-nucleus (A-A) collisions if the A-A 
collision is a simple superposition of N-N interactions. Consequently, $\Phi$ 
is independent of the centrality of A-A collision in such a case. On the other 
hand, $\Phi$ equals zero when the inter-particle correlations are entirely 
absent. The critical analysis of the $\Phi-$measure can be found in 
\cite{Tra00,Uty01}. For the central Pb-Pb collisions, the value of $\Phi(p_T)$ 
measured in the laboratory rapidity window (4.0, 5.5) equals $4.6 \pm 1.5$ 
MeV \cite{App99}. The preliminary result for proton-proton interactions in 
the same acceptance is $5 \pm 1$ MeV \cite{App99}. Although the two values 
are close to each other the mechanisms behind them seem to be very different. 
It has been shown \cite{App99} that the correlations, which are of the short 
range in the momentum space as those due to the Bose-Einstein statistics, are 
responsible for the positive value of $\Phi(p_T)$ in the central Pb-Pb collisions. 
When the short range correlations are excluded $\Phi(p_T)$ is reduced to 
$0.6 \pm 1$ MeV \cite{App99} Our calculations have indeed demonstrated 
\cite{Mro98,Mro99} that the effect of Bose statistics of pions modified by 
the hadron resonances fully explains the observed $\Phi(p_T)$ in the central 
Pb-Pb collisions. On the other hand, the short range correlations have been 
experimentally shown \cite{App99} to provide a negligible contribution to 
the $p_T-$fluctuations in the p-p interactions. Thus, the data suggest 
that the dynamical long range correlations are reduced in the central Pb-Pb 
collisions (when compared to p-p) with the short range due to the Bose 
statistics being amplified. The former feature is a natural consequence of 
the system evolution towards the thermodynamic equilibrium. The amplification 
of the quantum statistics effect results from the increased particle 
population in the final state phase-space. Since various dynamical 
correlations contribute to $\Phi(p_T)$ the question emerges what is the 
dynamical correlation in the nucleon-nucleon interactions which appear to be 
washed out in the central nucleus-nucleus collisions. The aim of this paper is 
to discuss the question. 

The average transverse momentum $\langle p_T \rangle$, which 
is measured at a given multiplicity $N$, is known to depend on $N$ in 
proton-proton collisions \cite{Aiv88}. The correlation is negative for 
the collision energies below, say, $\sqrt{s}=50$ GeV and positive for the
higher energies. At the beam energy of 205 GeV ($\sqrt{s}=19.7$ GeV),
which is very close to that of the NA49 measurement \cite{App99}, 
$\langle p_T \rangle$ significantly decreases 
with $N$ \cite{Kaf77}. The data are shown in Fig.~1, where we have 
merged the results for positive and negative particles. As already discussed 
in \cite{Gaz92}, the correlation which couples $\langle p_T \rangle$ 
to $N$ leads to $\Phi(p_T) > 0$. Here, we compute $\Phi(p_T)$ as a function 
of the correlation strength. Analytical and numerical results are presented. 
The effect of the finite acceptance is studied and comparison with the 
experimental data is performed.

\begin{figure}
\centerline{\epsfig{file=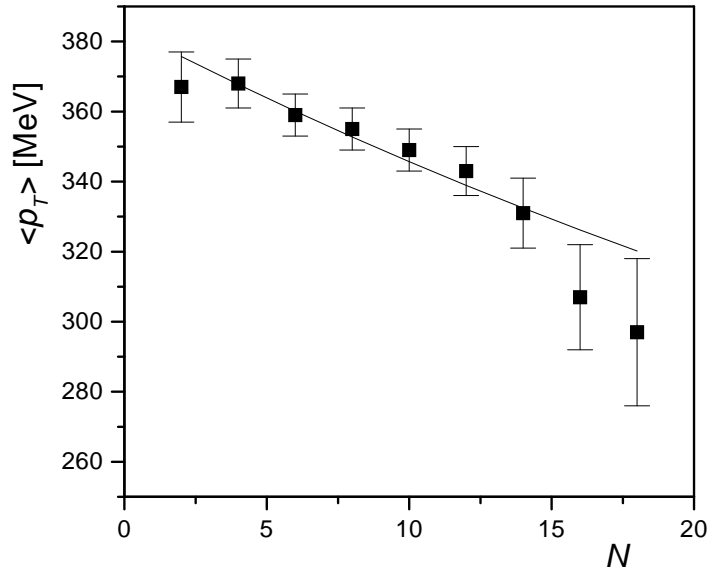,width=0.6\linewidth}}
\caption{ The average transverse momentum as a function of charged 
particle multiplicity in non-diffractive p-p collisions at 205 GeV. The 
data are taken from Ref. [8]. The line corresponds to 
$\langle N \rangle=6.56$, $T = 167$ MeV and 
$\Delta \! T = 1.25$ MeV, see the text for the parameter's meaning. }
\end{figure}

\section{Analytical calculation}

Let us first introduce the $\Phi-$measure. One defines the single-particle 
variable $z \buildrel \rm def \over = x - \overline{x}$ with the overline 
denoting averaging over a single particle inclusive distribution. Here, we 
identify $x$ with the particle transverse momentum. The event variable $Z$, 
which is a multiparticle analog of $z$, is defined as 
$Z \buildrel \rm def \over = \sum_{i=1}^{N}(x_i - \overline{x})$, where 
the summation runs over particles from a given event. By construction, 
$\langle Z \rangle = 0$ where $\langle ... \rangle$ represents averaging 
over events. Finally, the $\Phi-$measure is defined in the following way
$$
\Phi \buildrel \rm def \over = 
\sqrt{\langle Z^2 \rangle \over \langle N \rangle} -
\sqrt{\overline{z^2}} \;.
$$

The correlation $\langle p_T \rangle$ {\it vs.} $N$ is introduced to
our model calculations through the multiplicity dependent temperature 
or slope parameter of  $p_T-$distribution. Specifically, the single 
particle transverse momentum distribution in the events of multiplicity 
$N$ is chosen in the form suggested by the thermal model i.e.
\begin{equation}\label{pt-dis}
P_{(N)}(p_T) \sim p_T
\exp \bigg[ - {\sqrt{m^2 + p_T^2} \over T_N} \bigg] \;,
\end{equation}
where $m$ is the particle mass while $T_N$ is the multiplicity 
dependent temperature defined as
$$
T_N = T + \Delta \! T ( \langle N \rangle - N )
$$
with $\Delta \! T$ controlling the correlation strength. 
For $N=\langle N \rangle$ one gets $T_N = T$.  The inclusive 
transverse momentum distribution, which determines 
$\overline{z^2}= \overline{p^2_T} -\overline{p_T}^2$, reads 
$$
P_{\rm incl}(p_T) = {1 \over \langle N \rangle} 
\sum_N {\cal P}_N N P_{(N)}(p_T) \;,
$$
where ${\cal P}_N$ is the multiplicity distribution. 

The $N-$particle transverse momentum distribution in the events of 
multiplicity $N$ is assumed to be the $N-$product of $P_{(N)}(p_T)$. 
Therefore, all inter-particle correlations different than 
$\langle p_T \rangle$ {\it vs.} $N$ are entirely neglected. 
Then, one easily finds 
\begin{eqnarray*}
\langle Z^2 \rangle = \sum_N {\cal P}_N
\int_0^{\infty} dp_T^1 .\; . \; . \int_0^{\infty} dp_T^N 
\Big(p_T^1 +  .\; . \; . + p_T^N - N \overline{p_T} \Big)^2  
P_{(N)}(p_T^1) \; . \; . \; . \; P_{(N)}(p_T^N) \;.
\end{eqnarray*}

Assuming that the particles are massless and the correlation is weak
i.e. $T \gg \Delta \! T ( \langle N^2 \rangle - \langle N \rangle^2 )^{1/2}$
the calculation of $\Phi$ can be performed analytically. The results read:
\begin{eqnarray}\label{massless-phi}
\nonumber
{\langle Z^2 \rangle \over \langle N \rangle} &=& 2 T^2 
- 4 {T \Delta \! T \over \langle N \rangle}
\big(\langle N^2 \rangle - \langle N \rangle^2 \big) \\ \nonumber
&& \;\;\;\;\;\;\;
+ 2{\Delta \! T^2 \over \langle N \rangle^3} 
\big( 2\langle N^4 \rangle \langle N \rangle^2 
   - 4 \langle N^3 \rangle \langle N^2 \rangle \langle N \rangle
   +   \langle N^3 \rangle \langle N \rangle^2
   - 2 \langle N^2 \rangle \langle N \rangle^3 
   + 2 \langle N^2 \rangle^3
   +   \langle N \rangle^5 \big) \;,
\\ [3mm] \nonumber
\overline{z^2} &=& 2 T^2 
- 4 {T \Delta \! T \over \langle N \rangle} 
\big(\langle N^2 \rangle - \langle N \rangle^2 \big) \\ \nonumber
&& \;\;\;\;\;\;\;
+ 2{\Delta \! T^2 \over \langle N \rangle^2} 
\big(3 \langle N^3 \rangle \langle N \rangle
    -2 \langle N^2 \rangle \langle N \rangle^2
    +  \langle N \rangle^4 
    -2 \langle N^2 \rangle^2 \big) \;,
\\[3mm]
\Phi(p_T) &=& \sqrt{2}\, {\Delta \! T^2 \over T \langle N \rangle^3} 
\big( \langle N^4 \rangle \langle N \rangle^2
   -2 \langle N^3 \rangle \langle N^2\rangle \langle N \rangle
   -  \langle N^3 \rangle \langle N \rangle^2
    + \langle N^2 \rangle^3 
    + \langle N^2 \rangle^2 \langle N \rangle \big) \;, 
\end{eqnarray}
where terms of the third and higher powers of $\Delta \! T$ have 
been neglected. One observes that the lowest non-vanishing 
contribution to $\Phi$ is of the second order in $\Delta \! T$.
The above formulas are much simplified for the Poisson 
multiplicity distribution. Then, one finds
\begin{eqnarray*}
{\langle Z^2 \rangle \over \langle N \rangle} &=& 2 T^2 
- 4 T \Delta \! T 
+ 2 \Delta \! T^2 \big( 2\langle N \rangle^2 
+ 5\langle N \rangle + 1 \big)\;, 
\\ [3mm]
\overline{z^2} &=& 2 T^2 - 4 T \Delta \! T 
+ 2\Delta \! T^2 \big(3 \langle N \rangle + 1 \big)\;, 
\\[3mm]
\Phi(p_T) &=& \sqrt{2} \, {\Delta \! T^2 \over T } 
\big( \langle N \rangle^2 + \langle N \rangle \big) \;.
\end{eqnarray*}

The multiplicity distribution of charged particles  produced in high 
energy proton-proton collisions is, of course, not poissonian. First 
of all, the number of charged particles is always even due to the charge 
conservation. The multiplicity distribution of positive
(or negative) particles is also not poissonian - the dispersion does 
not grow as $\sqrt{\langle N \rangle}$ but it follows the so-called 
Wr\'oblewski formula \cite{Wro73} {\it i.e.} the dispersion is the 
linear function of $\langle N \rangle$. However, for the average 
multiplicities as low as those discussed here the Poisson distribution
provides a reasonable approximation. Therefore, the multiplicity  
distribution, which is further used in our calculation, is poissonian   
for negative pions with the number of positive pions being
exactly equal to that of negative charge. Thus, the charge conservation 
is satisfied in every event. For such a multiplicity distribution 
$\Phi (p_T)$, which is given by approximate Eq.~(\ref{massless-phi}), 
reads
\begin{equation}\label{analytic-phi}
\Phi(p_T) = 2 \sqrt{2} \, {\Delta \! T^2 \over T } 
\big( \langle N \rangle^2 + 3 \langle N \rangle \big) \;.
\end{equation}

\section{Numerical simulation}

The results from the previous section are instructive but 
the adopted approximations are very rough. So, let us present 
more realistic Monte Carlo calculations which can be confronted 
with the experimental data \cite{App99}. The proton-proton 
collisions are simulated event by event in the following way. For 
every event we first generate the multiplicity of negative particles 
from the Poisson distribution and then add the equal number of 
positive particles. The average multiplicity of negative particles 
has been taken as $\langle N^- \rangle =  3.28$ which is the 
experimentally observed negative multiplicity in non-diffractive 
proton-proton interactions at 205 GeV \cite{Bar74}.  This is the
collision energy corresponding to the data from Fig.~1. Further, 
we attribute the transverse momentum from the distribution
(\ref{pt-dis}) to each particle assuming that all particles are pions. 
The numerical values of  the temperature and correlation strength 
have been found fitting the data \cite{Kaf77} shown in Fig.~1. 
We have got  $T= 167 \pm 1.5$ MeV and  $\Delta \! T =  1.25 \pm 0.25 $ 
for charged particle multiplicity 
$\langle N \rangle =  2\langle N^- \rangle =  6.56$. 

Due to the particle registration inefficiency and finite detector coverage 
of the final state phase-space, only a fraction of  particles produced in 
high-energy collisions is usually observed the experimental studies.
As an input to our model calculations we have taken the 
$\langle p_T \rangle\!-\!N$ correlation observed 
in the full phase-space. The rapidity, which determines the acceptance 
domain, is not considered at all. Therefore, there is no difference 
whether a particle is lost due to the limited acceptance or due to the 
tracking inefficiency. Our Monte Carlo simulation takes into account 
the two effects in such a way that each generated particle - positive or 
negative pion - is registered with the probability $p$ and rejected with 
$(1-p)$. Below, we further discuss the procedure in the context of 
NA49 data \cite{App99}.  The results of our simulation are collected in 
Fig.~2 where $\Phi (p_T)$ as a function of $\Delta \! T$ for several values 
of $p$ is shown. We have checked that our simulation fully reproduces the 
results from \cite{Gaz92}, where only negative particles in the full 
acceptance have been studied.

\begin{figure}
\centerline{\epsfig{file=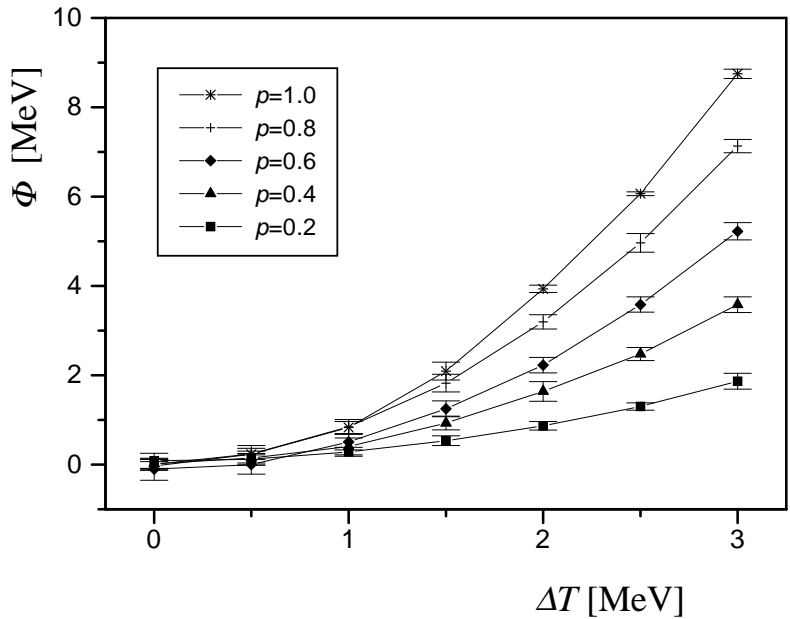,width=0.7\linewidth}}
\caption{$\Phi (p_T)$ as a function of the correlation strength
$\Delta \! T$ for several values of the acceptance probability $p$.}
\end{figure}

As seen in Fig.~2, $\Phi (p_T)$ grows quadratically with $\Delta \! T$ 
in agreement with the approximate equation (\ref{analytic-phi}). The 
increase of  $\Phi (p_T)$ with $p$ also follows from Eq.~(\ref{analytic-phi}). 
Indeed, when a particle is detected with the probability $p$ the temperature 
dispersion effectively decreases and the multiplicity $\langle N \rangle$ 
should be replaced by $p\langle N \rangle$ in (\ref{analytic-phi}). 
Consequently, $\Phi (p_T)$ grows quadratically with the observed particle
multiplicity and the correlation, which is easily observable in the full
phase-space, is hardly seen in the acceptance as small as 20\%. This
behavior is very different than that found in Refs. \cite{Mro98,Mro99}
where $\Phi (p_T)$ due to the Bose-Einstein correlations has been
computed. Then, $\Phi (p_T)$ is independent of the particle multiplicity.

The NA49 measurement of $\Phi (p_T)$ in proton-proton collisions 
has been performed in the transverse momentum and pion rapidity 
intervals (0.005, 1.5) GeV and (4.0, 5.5), respectively \cite{App99}. 
Only about 20\% of all produced particles have been observed. Our 
simulation gives $\Phi (p_T) =  0.41 \pm 0.07$ MeV for the values 
of  $\Delta \! T = 1.25$ MeV and $p=0.2$ which are adequate for 
the NA49 measurement. Thus, the theoretical result significantly 
underestimates the preliminary experimental one which is, as already noted, 
$\Phi (p_T) =  5 \pm 1$ MeV \cite{App99}. 

One wonders whether the discrepancy is not caused by our highly 
simplified procedure of taking into account the effect of finite acceptance. 
We first note that the rapidity coverage of  the NA49 measurement 
which is (4.0, 5.5) in the laboratory translates into (1.1, 2.6) in the 
center of mass frame. Therefore, the observed pions are not far the 
very central rapidity region. Since most of pions originates from 
the domain the average characteristics of all pions and that of the 
central ones are expected to be similar to each other. Indeed, the data 
from \cite{Kaf77} show that $\langle p_T \rangle$ for all pions and 
those from the central region are essentially the same. Therefore, 
the correlation strength parameter $\Delta \!T = 1.25$ MeV, which 
corresponds to the $\langle p_T \rangle\!-\!N$ correlation averaged 
over full phase-space, seem to be applicable not only to all pions but to 
the central ones as well. One further notes that $\langle p_T \rangle$ 
shown in \cite{Kaf77} changes with the rapidity similarly 
for different $N$. Consequently, the correlation 
$\langle p_T \rangle$ {\it vs.} $N$ only weakly varies 
with $y$ and it is hard to expect that the correlation strength 
observed in the NA49 acceptance domain is significantly larger than 
that averaged over the whole phase-space. The data \cite{Kaf77} 
suggests rather the opposite effect. Therefore, we conclude that  our 
simplified procedure cannot  distort the results dramatically and that 
the correlation $\langle p_T \rangle$ {\it vs.} $N$ does not 
explain the NA49 p-p preliminary data.

As already mentioned, $\Phi$ is constructed is such a way that it is exactly 
the same for nucleon-nucleon and nucleus-nucleus collisions if the latter 
is a simple superposition of  former ones. Therefore, $\Phi (p_T)$ shown 
in Fig.~2 holds for nucleus-nucleus when all secondary interactions are
neglected. Since our model neglects the Bose-Einstein correlations 
it can be compared with the NA49 data for the central Pb-Pb collisions
when the short range correlations are excluded. In such a case, our result 
$\Phi (p_T) = 0.41 \pm 0.07$ MeV for $\Delta \! T = 1.25$ MeV and 
$p=0.2$ appears to be compatible with the experimental value 
$\Phi (p_T)= 0.6 \pm 1$ MeV  \cite{App99}. The smallness of 
$\Phi (p_T)$ reported by NA49 collaboration is then not surprising 
at all. As follows from Fig.~2 it is caused by the limited acceptance. 
However, it is premature to draw a conclusion about $p_T-$correlations
in Pb-Pb collisions until the origin of $\Phi (p_T)$ p-p interactions 
is not explained.

\section{Concluding remarks}

We have studied how the correlation, which couples the average $p_T$ 
to the event multiplicity, influences the transverse momentum fluctuations 
observed by means of the $\Phi-$measure. The approximate analytical formula 
has been derived and then the numerical simulation has been performed. The 
effect of the finite detector acceptance has been taken into account. The 
procedure is highly simplified but it seems to be adequate for the NA49 data. 
It has been shown that the effect of the correlation 
$\langle p_T \rangle$ {\it vs.} $N$ is very weak if the particles from 
the small acceptance region are studied. Consequently, the correlation is far 
too weak to explain the preliminary experimental value of  $\Phi (p_T)$ in 
proton-proton collisions \cite{App99}. If the preliminary data is confirmed 
by the final analysis one should look for other sources of  dynamical 
$p_T-$fluctuations. The effect of the conservation laws presumably plays 
no role in the acceptance as small as 20\%.  The decays of hadron resonances 
which strongly correlate the momenta of decay products might be important. 
However, the hadron resonances are present in proton-proton and 
nucleus-nucleus collisions as well. Consequently, the reported reduction of 
long range dynamical correlations in the central Pb-Pb collisions when 
compared to p-p \cite{App99} remains unexplained. The situation is much 
simpler if preliminary data on $\Phi (p_T)$ in proton-proton collision 
\cite{App99} overestimates the real value. Then, $\Phi (p_T)$ from 
p-p and central Pb-Pb are close to each other. There is also no conflict 
between our calculations and the experimental data. However, one 
cannot conclude that the long range correlations present in the proton-proton 
interactions are washed out in the central heavy-ion collisions The problem 
obviously needs further experimental and theoretical studies. In particular, 
the data on proton-proton and nucleus-nucleus collisions from the enlarged 
acceptance are needed.

\begin{acknowledgements}

We are very grateful to Marek Ga\' zdzicki, Katarzyna Perl, Waldemar Retyk, 
and Ewa Skrzypczak for fruitful discussions and stimulating criticism.

\end{acknowledgements}

%\newpage


\vspace{-0.5cm}

\begin{thebibliography}{99}

\bibitem{App99} H.~Appelsh\"auser {\it et al.}, 
Phys. Lett. B {\bf 459}, 679 (1999).

\bibitem{Gaz92} M.~Ga\' zdzicki and St.~Mr\' owczy\' nski, 
Z. Phys. C {\bf 54}, 127 (1992).

\bibitem{Tra00} T.A.~Trainor, hep-ph/0001148.

\bibitem{Uty01} O.V.~Utyuzh, G.~Wilk, and Z.~W\l odarczyk, hep-ph/0103158,
to appear in Phys. Rev. C.

\bibitem{Mro98} St.~Mr\'owczy\'nski, Phys. Lett. B {\bf 439}, 6 (1998).

\bibitem{Mro99} St.~Mr\'owczy\'nski, Phys. Lett. B {\bf 465}, 8 (1999).

\bibitem{Aiv88} V.V.~Aivazyan {\it et al.}, Phys. Lett. B {\bf 209}, 103 (1988).

\bibitem{Kaf77} T.~Kafka {\it et al.}, Phys. Rev. D {\bf 16}, 1261 (1977).

\bibitem{Bar74} S.~Barish {\it et al.}, Phys. Rev. D {\bf 9}, 2689 (1974).

\bibitem{Wro73} A.~Wr\'oblewski, Acta Phys. Pol. B {\bf 4}, 857 (1973). 

\end{thebibliography}
\end{document}